\documentclass[prl,aps,showpacs,epsf, twocolumn]{revtex4}

\def\be{\begin{equation}}
\def\ee{\end{equation}}

\usepackage{graphicx}
\usepackage{bm}

\begin{document}

\title{Collective Oscillations of an Imbalanced Fermi Gas: \\Axial Compression Modes and Polaron Effective Mass}
\author{S. Nascimb\`ene$^1$, N. Navon$^1$, K. J. Jiang$^1$, L. Tarruell$^2$, M. Teichmann$^3$, J. McKeever$^4$, F. Chevy$^1$, and C. Salomon$^1$}

\affiliation{$^1$Laboratoire Kastler Brossel, CNRS, UPMC,
\'Ecole Normale Sup\'erieure, 24 rue Lhomond, 75231 Paris, France\\
$^2$Institute for Quantum Electronics, ETH Zurich, 8093 Zurich, Switzerland\\
$^3$Max-Born-Institut, Max-Born-Strasse 2 A, 12489 Berlin, Germany \\
$^4$IOS, CQIQC and Dept. of Physics, University of Toronto, Canada}

\date{\today}

\begin{abstract}
We investigate the low-lying compression modes of a unitary Fermi gas with imbalanced spin populations.
For low polarization, the strong coupling between the two spin components leads to a hydrodynamic behavior of the cloud.
For large population imbalance we observe a decoupling of the oscillations of the two spin components,
giving access to the effective mass of the Fermi polaron, a quasi-particle composed of an
impurity dressed by particle-hole pair excitations in a surrounding Fermi sea.  We find $m^*/m=1.17(10)$, in agreement
with the most recent theoretical predictions.
\end{abstract}

 \maketitle

The study of the low lying excitation modes of a complex system can
be a powerful tool for investigation of its physical properties.  For instance, the Earth's structure has been probed using the
propagation of seismic waves in the mantle, and the ripples in
space-time propagated by gravitational waves can be used as
probes of extreme cosmic phenomena. In ultra-cold atomic gases, the
measurement of low energy modes of bosonic or fermionic systems has
been used to probe superfluidity effects \cite{giorgini2008theory}, to measure the
angular momentum of vortex lattices \cite{Chevy2000MAM} and to
characterize the equation of state of fermionic superfluids
\cite{bartenstein2004cmo,kinast2004esr}.

\begin{figure}
\centerline{\includegraphics[width=0.93\columnwidth]{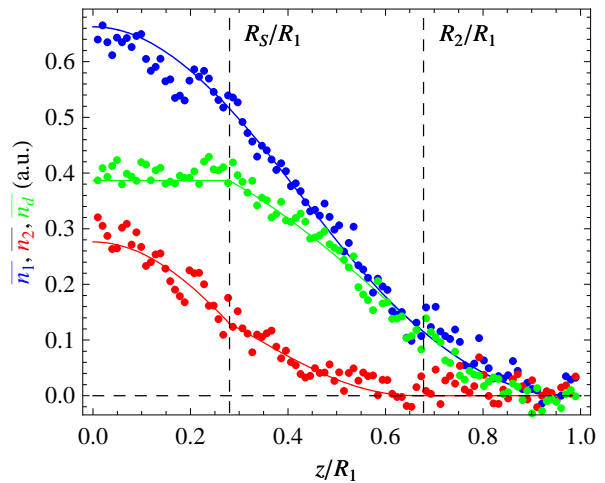}}
\caption{(color online) Integrated density profiles of an imbalanced Fermi gas.
Blue: majority atoms $\bar n_1(z)$; Red: minority atoms $\bar
n_2(z)$; Green: difference $\bar n_d=\bar n_1-\bar n_2$. In this
latter case, the flat-top feature signals a cancellation of the
density difference at the center of the trap, characteristic of the
existence of a fully paired superfluid core. The superfluid (resp. minority) radius $R_S$ (resp. $R_2$) are
marked by vertical dashed lines. The full color lines
correspond to the prediction of Monte-Carlo theories
\cite{recati2008ris}, the only fit parameters being the number of
atoms in each spin state, $N_1=8.0\times10^4,\;N_2=2.4\times10^4$
for this image. The axial (radial) trap frequency is 18.6 Hz (420
Hz).} \label{Fig1}
\end{figure}

In this paper, we study the excitation spectrum of an ultra-cold
Fermi gas with imbalanced spin populations. This topic was
initiated in the 60's by the seminal works of Clogston and
Chandrasekhar \cite{clogston1962ulc,Chandrasekhar1962} and
found only recently an experimental confirmation thanks to the
latest developments in ultra-cold Fermi gases
\cite{partridge2006pap,zwierlein2006fsi}. These dramatic experiments
have observed that when a fermionic superfluid is polarized through imbalance of spin populations, the trapped atomic cloud forms a
shell structure. The energy gap associated with pairing maintains a superfluid core where the two spin densities
are equal, while the outer shell is composed by a normal gas with imbalanced spin densities
(see Fig.\ref{Fig1}).
Here, we extend this work to the dynamical properties of these
systems and we focus on the regime of strong interactions, where the scattering length $a$ is infinite.
We show in particular that the study of the axial breathing
mode provides valuable insight on the dynamical properties of a
quasi-particle, the Fermi polaron, that was introduced recently
to describe the normal component occupying the outer shell of the
cloud
\cite{lobo2006nsp,chevy2006upa,combescot2008nsh,combescot2007nsh,prokofev2008bdm,schirotzek2009ofp}.
The Fermi polaron is composed of an impurity (labelled $2$) immersed in a
non-interacting Fermi sea (labelled $1$), and is analogous to the polaron of condensed
matter physics, {\it i.e.} an electron immersed in a bath of
non-interacting (bosonic) phonons. According to the Landau theory of the
Fermi liquid, the low energy spectrum of the polaron is similar to
that of a free particle and can, in the local density approximation (LDA),
be recast as \be
  E_2(\bm r,\bm p) =  A E_{F1}(\bm r)+V(\bm r)+\frac{p^2}{2m^*}+...\label{eqEnergyPolaron}
\ee

\noindent where $V$ is the trapping potential, $E_{F1}(\bm
r)=E_{F1}(\bm 0)-V(\bm r)$ is the local Fermi energy of the majority
species, $A$ is a dimensionless quantity characterizing the attraction 
of the impurity by the majority atoms and $m^*$ is the effective mass
of the Fermi polaron. For $a=\infty$, $A=-0.61$ has been determined both
experimentally \cite{schirotzek2009ofp} and theoretically
\cite{prokofev2008bdm,combescot2008nsh,combescot2007nsh,lobo2006nsp,chevy2006upa},
while slight disagreements still exist on the value of the effective
mass. Fixed node Monte-Carlo suggests $m^*/m=1.09(2)$
\cite{pilati2008psi}, systematic diagrammatic expansion yields
$m^*/m=1.20$ \cite{combescot2008nsh} and analysis of density
profiles (such as Fig.\ref{Fig1}) gives $m^*/m=1.06$ \cite{shin2008des}.

From Eq. (\ref{eqEnergyPolaron}), the quasi-particle evolves in an effective potential
$V^*(\bm r)=(1-A)V(\bm r)$. Assuming $V(\bm r)$ to be harmonic with frequency $\omega$, the
polaron is trapped in an effective potential of frequency $\omega^*$ \cite{lobo2006nsp}:
\be
\frac{\omega^*}{\omega}=\sqrt{\frac{1-A}{m^*/m}}.\label{EqnFrequency}
\ee In this paper we determine the effective mass through the
measurement of the oscillation frequency $\omega^*$ in the axial
direction (labelled $z$) of a cylindrically symmetric trap.

Our experimental setup is an upgraded version of the one presented
in \cite{bourdel2004esb}. $7\times10^6$~$^6$Li atoms in the hyperfine state $|F=3/2,
m_F=+3/2\rangle$ are loaded into a mixed magnetic/optical trap at
$100~\mu$K. The optical trap uses a single beam of waist
$w_0=35~\mu$m and maximum power $P=60$~W operating at a wavelength
$\lambda=1073$~nm. The atoms are transferred into the
hyperfine ground state $|1/2, 1/2\rangle$, and a spin mixture
is created by a radio-frequency sweep across the hyperfine
transition $|1/2,1/2\rangle\rightarrow |1/2,
-1/2\rangle$. By varying the rate of this sweep, we control the sample's degree of polarization $P\equiv(N_1-N_2)/(N_1+N_2)$, where $N_1$ (resp. $N_2$)
is the atom number of the majority (resp. minority) spin species. The mixture is then evaporatively cooled in $6~\mathrm{s}$ by reducing the laser power to $70$~mW.
This is done at a magnetic field $B=834$~G, which
corresponds to the position of the broad Feshbach resonance in $^6$Li
where the scattering length is infinite and where further
experiments are performed. Typical radial frequencies are $\omega_x=\omega_y\sim2\pi\times400$ Hz.
The axial confinement of the dipole trap
is enhanced by the addition of a magnetic curvature, leading to an
axial frequency $\omega_z\sim2\pi\times30$~Hz. Our samples contain $\sim 8\times10^4$ atoms in the
majority spin state at a temperature $T\lesssim 0.09\;T_F$.
The temperature is evaluated by fitting the wings of the majority
density profile outside the minority radius. In this region, the gas
is non-interacting, allowing unambiguous thermometry of the inner,
strongly-interacting part of the cloud \cite{shin2008pd}. Here,
$T_F$ is defined as the Fermi temperature of an ideal gas whose
density profile overlaps the majority one in the fully polarized
rim. Our thermometry's precision is limited by the finite signal
to noise ratio of the data, hence the quoted upper bound.

The two spin states are imaged sequentially using {\em in situ}
absorption imaging. To prevent heating from the scattered photons
and the strong interactions between the two species, the duration of
the two imaging pulses as well as their separation must be short
(10~$\mu$s each in our case). By reversing the order in which we
image the two spin components, we checked that the imaging of the
first species did not significantly influence the second. Typical
integrated density profiles of the atom cloud $\bar n(z)=\int
\textrm{d}x\textrm{d}y\;n(x,y,z)$, where $n(x,y,z)$ is the 3D atom
density, are presented in Fig.\ref{Fig1}. These profiles display the
characteristic features already observed by the MIT group
\cite{shin2008pd}: a flat-top structure in the superfluid region
confirming the existence of a fully paired core satisfying the LDA \cite{desilva2006stu}, an intermediate phase
where the two spin species are present with unequal densities, and an
outer rim containing only majority atoms. Following \cite{recati2008ris}, we compare our density profiles to the
prediction for the equation of state of the different phases and
find fairly good agreement. In particular, we observe that the superfluid core disappears for
polarizations $P>0.76(3)$. This limit agrees well with the measurement of the MIT group \cite{zwierlein2006fsi} and differs
from the Rice group value \cite{partridge2006pap}.

We excite the axial breathing mode by switching off the
axial magnetic trapping field for 1~ms. The effect of this excitation is
twofold: in addition to nearly suppressing the axial
confinement, the bias field is increased up to 1050~G, where $k_F a\sim -1$,
so that the gas is no longer strongly interacting. This scheme provides a spatially
selective excitation of the cloud. Indeed, while the reduction of
the trapping frequency perturbs the whole cloud, the modification of
the scattering length only acts in the region where the two spin
components overlap. In the regime of strong polarization, these two
regions are well separated, leading to a differential excitation of the two spin components.

\begin{figure}
\vspace{2mm}
\centerline{\hspace{-1.5cm}\includegraphics[width=0.55\columnwidth]{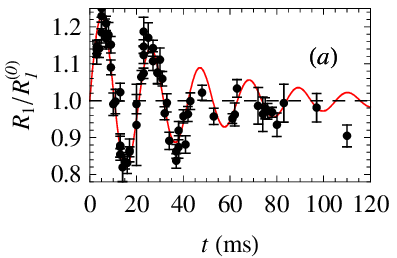}}
\vspace{-3.5cm}
\centerline{\includegraphics[width=0.95\columnwidth]{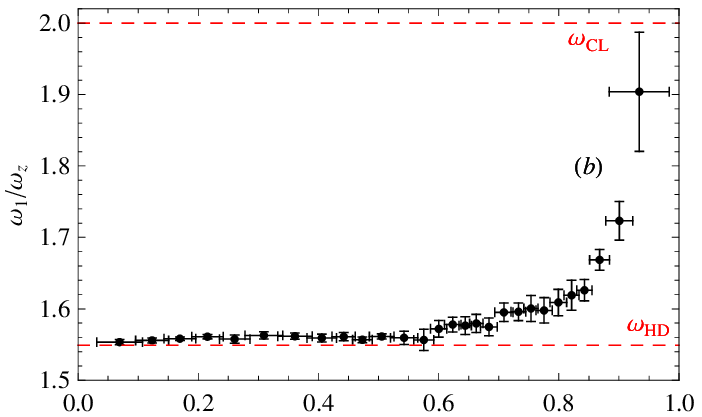}}
\centerline{\hspace{-0.35cm}\includegraphics[width=0.99\columnwidth]{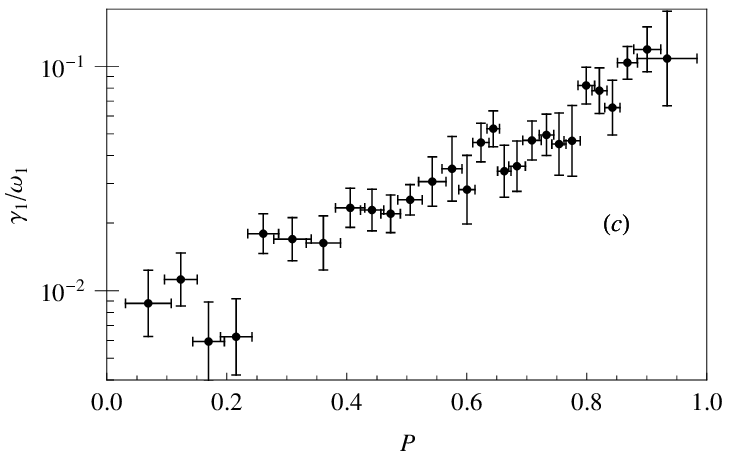}}
\caption{(a) Oscillations of the axial radius of
the majority component, for a population imbalance $P=0.85(2)$,
beyond the Clogston limit. The full line corresponds to a fit by an
exponentially damped sinusoid. (b) Frequency of the breathing mode
$\omega_1$ normalized to the axial trapping frequency $\omega_z$.
The superfluid (resp. collisionless) limits $\omega_1
=\sqrt{12/5}\,\omega_z$ (resp. $2\,\omega_z$) are indicated by the
red lines. The axial (radial) trap frequency is 28.9(1) Hz (420 Hz).
(c) Damping rate $\gamma_1$ as the function of polarization (in log
scale). Note that our data are limited to $P<0.95$ due to the small
minority atom number ($N_2\lesssim2\times10^3$) at such high
polarizations.} \label{Fig2}
\end{figure}

Let us first focus on the oscillations of the majority spin species
presented in Fig.\ref{Fig2}. Typical dynamics of the outer radius
$R_1(t)$ of the majority component are exemplified by Fig.\ref{Fig2}a.
For each polarization, this time evolution is
fitted using an exponentially damped sinusoid, with
$R_1(t)=R_1^{(0)}(1+A_1\cos(\omega_1 t+\varphi) e^{-\gamma_1 t})$,
and the variations of $\omega_1$ and $\gamma_1$ as a function of $P$
are displayed in Fig.\ref{Fig2}b and Fig.\ref{Fig2}c. One remarkable feature of
this graph is the frequency plateau for polarizations
$P\lesssim0.7$, corresponding approximately to the domain where a
superfluid core is present in the cloud. Although in this range of
parameters, the dynamics of the system is fairly complex due to the
strong coupling between the superfluid and normal components, a
simple argument based on a sum rule approach generalizing the result
of \cite{vichi1999coi} allows us to understand this property.

We describe the system by the Hamiltonian $H=\sum_i p_i^2/2m+U(\bm
r_1,\bm r_2,...)$, where $\bm r_i$ (resp. $\bm p_i$) is the position
(resp. momentum of particle $i$), $m$ is the mass of the atoms and
$U$ includes both trapping potential and interatomic interaction.
The compression of the trapping frequency in the $z$ direction is
associated with the operator $F=\sum_i z_i^2$. Let us introduce
the moments of the spectral distribution associated with $F$ and
defined by
$$m_k=\sum_{n\neq0} (E_n-E_0)^k\left|\left\langle 0\left|F\right|n\right\rangle\right|^2,$$
\noindent where the $|n\rangle$ are the eigenstates of $H$
associated with the eigenvalue $E_n$, and $|0\rangle$ is the
many-body ground state. We assume that the operator
$F$ mainly couples $|0\rangle$ to one excited state $|1\rangle$. In
this case, the frequency of the breathing mode excited by the axial
compression of the trap is given by
$\omega_1=(E_1-E_0)/\hbar\simeq\sqrt{m_1/m_{-1}}/\hbar$. An explicit calculation of
these two moments leads to the following expression : \be
\omega_1^2\simeq-2\langle z^2\rangle\left/\frac{\partial \langle
z^2\rangle}{\partial\omega_z^2}\right.. \ee For a unitary gas, LDA imposes that the mean radius of the cloud is
given by $\langle z^2\rangle = R_{\rm TF}^2 f(P)$, where $R_{\rm
TF}$ is the radius of an ideal Fermi gas in the same trap and with
the same atom number and $f$ is some universal function of the
polarization \cite{chevy2006dpt}. Using this assumption, the
calculation of the oscillation frequency is straightforward and
yields $\sqrt{\frac{12}{5}}\omega_z=1.55\omega_z$, {\it i.e.} the hydrodynamic
prediction \cite{bartenstein2004cmo,amoruso1999cet} for $P=0$, {\em regardless of the polarization of the
sample}. This argument is in good agreement with our experimental
findings (Fig.\ref{Fig2}b).

\begin{figure}
\centerline{\includegraphics[width=0.9\columnwidth]{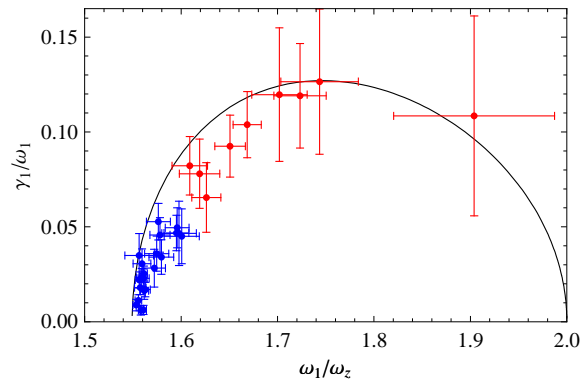}}
\caption{Comparison of our experimental results with the parametric curve
$(\omega_1(\tau)/\omega_z,\gamma_1(\tau)/\omega_1(\tau))$ deduced from prediction (\ref{Eqn2}).
The data in blue (red)
correspond to polarizations $P<0.8$ ($P>0.8$). } \label{Fig3}
\end{figure}

At larger polarizations the frequency sharply increases towards the collisionless value. The damping rate, very small in the balanced superfluid, increases by a
factor $\sim 20$ for higher imbalances \cite{dampingrate}. Interestingly, as seen in Fig.\ref{Fig3}, this behavior is consistent
with a general argument about relaxation processes in fluid dynamics \cite{landau1987fmv}.
Indeed, one can relate $\omega_1$ and $\gamma_1$ through

\be \omega^2=\omega_{\rm CL}^2+\frac{\omega_{\rm HD}^2-\omega_{\rm
CL}^2}{1+i\omega\tau}, \label{Eqn2} \ee

\noindent where $\omega=\omega_1+i\gamma_1$, $\omega_{\rm
HD}=\sqrt{12/5}\,\omega_z$ (resp. $\omega_{\rm CL}=2\,\omega_z$) is
the hydrodynamic (resp. collisionless) frequency and $\tau$ is an
effective relaxation rate.

Measurements of $\omega_1/\omega_z$ in three different traps of
aspect ratios 8.2, 9.0 and 14.5 give identical results (within 3\%) for all
polarizations. By contrast, the effect of temperature is more
pronounced.
At $0.12(1)\,T_{F}$, $\omega_1(P)$ remains equal to the hydrodynamic
prediction at all attainable polarizations with $P_{\text{max}}=0.95$
, for a cloud of $N_1\sim2\times10^5$ majority atoms held in a trap of aspect ratio 22. This illustrates the
role of Pauli blocking at the lowest temperatures which favors collisionless behavior.

\begin{figure}
\centerline{\hspace{3.0cm}\includegraphics[width=0.485\columnwidth]{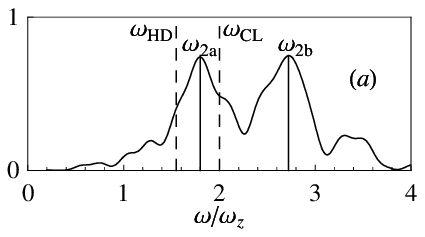}}
\vspace{-2.85cm}
\centerline{\includegraphics[width=1.06\columnwidth]{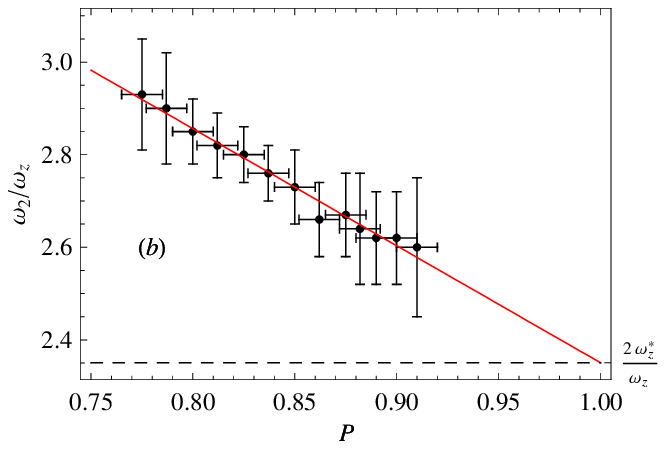}}
\caption{(a) Frequency power spectrum for $P=0.90(2)$.
The peak between $\omega_{\rm HD}$ and $\omega_{\rm CL}$ corresponds
to the oscillation in phase with the majority, the other one to the
polaron oscillation. (b) Frequency of the polaron component as a
function of polarization. All frequencies are normalized to $\omega_z$.}
\label{Fig4}
\end{figure}

Let us now consider the dynamics of the minority cloud (we recall that subscript $2$ refers to the impurity atoms). We observe
that for polarizations smaller than $P\sim 0.75$, the oscillation
frequencies and damping rates of the two spin species are equal, indicating a strong
coupling between them. By contrast, for $P>0.75$, a Fourier spectrum
of $R_2(t)$ reveals two frequencies (Fig.\ref{Fig4}a). The
lower frequency $\omega_{2a}$ is equal to the majority oscillation frequency $\omega_{1}$.
 We interpret the higher frequency $\omega_{2b}$, whose weight increases with
polarization, as the axial breathing of the minority atoms out of
phase with the majority cloud. A linear extrapolation of this
frequency to $P=1$ gives the oscillation frequency of a dilute gas
of weakly interacting polarons inside a Fermi sea at rest,
$\omega_{2b}(P\rightarrow1)=2.35(10)\omega_z$ (Fig.\ref{Fig4}b). The
uncertainty represents the standard deviation of a linear fit taking
into account the statistical uncertainties of the $\omega_{2b}$ measurements for each polarization.

By identifying the breathing mode frequency $\omega_{2b}$ as
$2\omega_z^*$ and using (\ref{EqnFrequency}), we deduce the mass of the quasi-particle : $m^*/m=1.17(10)$.
This is the first dynamic measurement of the polaron effective mass, in good agreement with the most recent theoretical predictions \cite{pilati2008psp,combescot2008nsh}, and with static observations \cite{shin2008des,recati2008ris}. $m^*$ is close to $m$ (albeit different), a surprising feature for a strongly interacting system at unitarity. We also exclude a mass $m^*=2m$ that one would expect for a deeply bound bosonic molecule.

In conclusion, we have studied the low frequency breathing modes of
an elongated Fermi gas with imbalanced spin populations. In the
presence of a superfluid core, the majority and minority components oscillate in phase
with a frequency that is largely independent of the spin polarization and in agreement with
the hydrodynamic prediction. At strong polarizations, the minority atom oscillation reveals
a second frequency, that we interpret as the Fermi polaron breathing mode.
Further investigations will extend our work to all
values of the scattering length. In particular, they should provide a clear
signature of the polaron-molecule transition \cite{prokof'ev08fpb,schirotzek2009ofp}. The role of interactions between polarons and damping phenomena should also be clarified \cite{bruun2008collisional}.

We are grateful to S. Stringari, A. Recati, C. Lobo, M. Zwierlein, J. Dalibard and Y. Castin for fruitful
discussions as well as K. Magalh\~aes and G. Duffy for
 experimental support. We acknowledge
support from ESF (FerMix), SCALA, ANR FABIOLA, R\'egion Ile de France
(IFRAF), ERC and Institut Universitaire de France.

\bibliographystyle{apsrev}
\bibliography{Imbalanced2}

\begin{thebibliography}{28}
\expandafter\ifx\csname natexlab\endcsname\relax\def\natexlab#1{#1}\fi
\expandafter\ifx\csname bibnamefont\endcsname\relax
  \def\bibnamefont#1{#1}\fi
\expandafter\ifx\csname bibfnamefont\endcsname\relax
  \def\bibfnamefont#1{#1}\fi
\expandafter\ifx\csname citenamefont\endcsname\relax
  \def\citenamefont#1{#1}\fi
\expandafter\ifx\csname url\endcsname\relax
  \def\url#1{\texttt{#1}}\fi
\expandafter\ifx\csname urlprefix\endcsname\relax\def\urlprefix{URL }\fi
\providecommand{\bibinfo}[2]{#2}
\providecommand{\eprint}[2][]{\url{#2}}

\bibitem[{\citenamefont{Giorgini et~al.}(2008)\citenamefont{Giorgini,
  Pitaevskii, and Stringari}}]{giorgini2008theory}
\bibinfo{author}{\bibfnamefont{S.}~\bibnamefont{Giorgini}},
  \bibinfo{author}{\bibfnamefont{L.}~\bibnamefont{Pitaevskii}},
  \bibnamefont{and}
  \bibinfo{author}{\bibfnamefont{S.}~\bibnamefont{Stringari}},
  \bibinfo{journal}{Rev. Mod. Phys.} \textbf{\bibinfo{volume}{80}},
  \bibinfo{pages}{1215} (\bibinfo{year}{2008}).

\bibitem[{\citenamefont{Chevy et~al.}(2000)\citenamefont{Chevy, Madison, and
  Dalibard}}]{Chevy2000MAM}
\bibinfo{author}{\bibfnamefont{F.}~\bibnamefont{Chevy}},
  \bibinfo{author}{\bibfnamefont{K.~W.} \bibnamefont{Madison}},
  \bibnamefont{and} \bibinfo{author}{\bibfnamefont{J.}~\bibnamefont{Dalibard}},
  \bibinfo{journal}{Phys. Rev. Lett.} \textbf{\bibinfo{volume}{85}},
  \bibinfo{pages}{2223} (\bibinfo{year}{2000}).

\bibitem[{\citenamefont{Bartenstein et~al.}(2004)\citenamefont{Bartenstein,
  Altmeyer, Riedl, Jochim, Chin, Denschlag, and Grimm}}]{bartenstein2004cmo}
\bibinfo{author}{\bibfnamefont{M.}~\bibnamefont{Bartenstein}},
  \bibinfo{author}{\bibfnamefont{A.}~\bibnamefont{Altmeyer}},
  \bibinfo{author}{\bibfnamefont{S.}~\bibnamefont{Riedl}},
  \bibinfo{author}{\bibfnamefont{S.}~\bibnamefont{Jochim}},
  \bibinfo{author}{\bibfnamefont{C.}~\bibnamefont{Chin}},
  \bibinfo{author}{\bibfnamefont{J.}~\bibnamefont{Denschlag}},
  \bibnamefont{and} \bibinfo{author}{\bibfnamefont{R.}~\bibnamefont{Grimm}},
  \bibinfo{journal}{Phys. Rev. Lett.} \textbf{\bibinfo{volume}{92}},
  \bibinfo{pages}{203201} (\bibinfo{year}{2004}).

\bibitem[{\citenamefont{Kinast et~al.}(2004)\citenamefont{Kinast, Hemmer, Gehm,
  Turlapov, and Thomas}}]{kinast2004esr}
\bibinfo{author}{\bibfnamefont{J.}~\bibnamefont{Kinast}},
  \bibinfo{author}{\bibfnamefont{S.}~\bibnamefont{Hemmer}},
  \bibinfo{author}{\bibfnamefont{M.}~\bibnamefont{Gehm}},
  \bibinfo{author}{\bibfnamefont{A.}~\bibnamefont{Turlapov}}, \bibnamefont{and}
  \bibinfo{author}{\bibfnamefont{J.}~\bibnamefont{Thomas}},
  \bibinfo{journal}{Phys. Rev. Lett.} \textbf{\bibinfo{volume}{92}},
  \bibinfo{pages}{150402} (\bibinfo{year}{2004}).

\bibitem[{\citenamefont{Clogston}(1962)}]{clogston1962ulc}
\bibinfo{author}{\bibfnamefont{A.}~\bibnamefont{Clogston}},
  \bibinfo{journal}{Phys. Rev. Lett.} \textbf{\bibinfo{volume}{9}},
  \bibinfo{pages}{266} (\bibinfo{year}{1962}).

\bibitem[{\citenamefont{Chandrasekhar}(1962)}]{Chandrasekhar1962}
\bibinfo{author}{\bibfnamefont{B.~S.} \bibnamefont{Chandrasekhar}},
  \bibinfo{journal}{App. Phys. Lett.} \textbf{\bibinfo{volume}{1}},
  \bibinfo{pages}{7} (\bibinfo{year}{1962}).

\bibitem[{\citenamefont{Partridge et~al.}(2006)\citenamefont{Partridge, Li,
  Kamar, Liao, and Hulet}}]{partridge2006pap}
\bibinfo{author}{\bibfnamefont{G.}~\bibnamefont{Partridge}},
  \bibinfo{author}{\bibfnamefont{W.}~\bibnamefont{Li}},
  \bibinfo{author}{\bibfnamefont{R.}~\bibnamefont{Kamar}},
  \bibinfo{author}{\bibfnamefont{Y.}~\bibnamefont{Liao}}, \bibnamefont{and}
  \bibinfo{author}{\bibfnamefont{R.}~\bibnamefont{Hulet}},
  \bibinfo{journal}{Science} \textbf{\bibinfo{volume}{311}},
  \bibinfo{pages}{503} (\bibinfo{year}{2006}).

\bibitem[{\citenamefont{Zwierlein et~al.}(2006)\citenamefont{Zwierlein,
  Schirotzek, Schunck, and Ketterle}}]{zwierlein2006fsi}
\bibinfo{author}{\bibfnamefont{M.}~\bibnamefont{Zwierlein}},
  \bibinfo{author}{\bibfnamefont{A.}~\bibnamefont{Schirotzek}},
  \bibinfo{author}{\bibfnamefont{C.}~\bibnamefont{Schunck}}, \bibnamefont{and}
  \bibinfo{author}{\bibfnamefont{W.}~\bibnamefont{Ketterle}},
  \bibinfo{journal}{Science} \textbf{\bibinfo{volume}{311}},
  \bibinfo{pages}{492} (\bibinfo{year}{2006}).

\bibitem[{\citenamefont{Lobo et~al.}(2006)\citenamefont{Lobo, Recati, Giorgini,
  and Stringari}}]{lobo2006nsp}
\bibinfo{author}{\bibfnamefont{C.}~\bibnamefont{Lobo}},
  \bibinfo{author}{\bibfnamefont{A.}~\bibnamefont{Recati}},
  \bibinfo{author}{\bibfnamefont{S.}~\bibnamefont{Giorgini}}, \bibnamefont{and}
  \bibinfo{author}{\bibfnamefont{S.}~\bibnamefont{Stringari}},
  \bibinfo{journal}{Phys. Rev. Lett.} \textbf{\bibinfo{volume}{97}},
  \bibinfo{pages}{200403} (\bibinfo{year}{2006}).

\bibitem[{\citenamefont{Chevy}(2006{\natexlab{a}})}]{chevy2006upa}
\bibinfo{author}{\bibfnamefont{F.}~\bibnamefont{Chevy}},
  \bibinfo{journal}{Phys. Rev. A} \textbf{\bibinfo{volume}{74}},
  \bibinfo{eid}{063628} (\bibinfo{year}{2006}{\natexlab{a}}).

\bibitem[{\citenamefont{Combescot and Giraud}(2008)}]{combescot2008nsh}
\bibinfo{author}{\bibfnamefont{R.}~\bibnamefont{Combescot}} \bibnamefont{and}
  \bibinfo{author}{\bibfnamefont{S.}~\bibnamefont{Giraud}},
  \bibinfo{journal}{Phys. Rev. Lett.} \textbf{\bibinfo{volume}{101}},
  \bibinfo{pages}{050404} (\bibinfo{year}{2008}).

\bibitem[{\citenamefont{Combescot et~al.}(2007)\citenamefont{Combescot, Recati,
  Lobo, and Chevy}}]{combescot2007nsh}
\bibinfo{author}{\bibfnamefont{R.}~\bibnamefont{Combescot}},
  \bibinfo{author}{\bibfnamefont{A.}~\bibnamefont{Recati}},
  \bibinfo{author}{\bibfnamefont{C.}~\bibnamefont{Lobo}}, \bibnamefont{and}
  \bibinfo{author}{\bibfnamefont{F.}~\bibnamefont{Chevy}},
  \bibinfo{journal}{Phys. Rev. Lett.} \textbf{\bibinfo{volume}{98}},
  \bibinfo{pages}{180402} (\bibinfo{year}{2007}).

\bibitem[{\citenamefont{Prokof'ev and
  Svistunov}(2008{\natexlab{a}})}]{prokofev2008bdm}
\bibinfo{author}{\bibfnamefont{N.}~\bibnamefont{Prokof'ev}} \bibnamefont{and}
  \bibinfo{author}{\bibfnamefont{B.}~\bibnamefont{Svistunov}},
  \bibinfo{journal}{Phys. Rev. B} \textbf{\bibinfo{volume}{77}},
  \bibinfo{pages}{125101} (\bibinfo{year}{2008}{\natexlab{a}}).

\bibitem[{\citenamefont{Schirotzek et~al.}(2009)\citenamefont{Schirotzek, Wu,
  Sommer, and Zwierlein}}]{schirotzek2009ofp}
\bibinfo{author}{\bibfnamefont{A.}~\bibnamefont{Schirotzek}},
  \bibinfo{author}{\bibfnamefont{C.-H.} \bibnamefont{Wu}},
  \bibinfo{author}{\bibfnamefont{A.}~\bibnamefont{Sommer}}, \bibnamefont{and}
  \bibinfo{author}{\bibfnamefont{M.~W.} \bibnamefont{Zwierlein}},
  \bibinfo{journal}{Phys. Rev. Lett.} \textbf{\bibinfo{volume}{102}},
  \bibinfo{eid}{230402} (\bibinfo{year}{2009}).

\bibitem[{\citenamefont{Pilati and
  Giorgini}(2008{\natexlab{a}})}]{pilati2008psi}
\bibinfo{author}{\bibfnamefont{S.}~\bibnamefont{Pilati}} \bibnamefont{and}
  \bibinfo{author}{\bibfnamefont{S.}~\bibnamefont{Giorgini}},
  \bibinfo{journal}{Phys. Rev. Lett.} \textbf{\bibinfo{volume}{100}},
  \bibinfo{pages}{030401} (\bibinfo{year}{2008}{\natexlab{a}}).

\bibitem[{\citenamefont{Shin}(2008)}]{shin2008des}
\bibinfo{author}{\bibfnamefont{Y.}~\bibnamefont{Shin}}, \bibinfo{journal}{Phys.
  Rev. A} \textbf{\bibinfo{volume}{77}}, \bibinfo{eid}{041603}
  (\bibinfo{year}{2008}).

\bibitem[{\citenamefont{Bourdel et~al.}(2004)\citenamefont{Bourdel, Khaykovich,
  Cubizolles, Zhang, Chevy, Teichmann, Tarruell, Kokkelmans, and
  Salomon}}]{bourdel2004esb}
\bibinfo{author}{\bibfnamefont{T.}~\bibnamefont{Bourdel}},
  \bibinfo{author}{\bibfnamefont{L.}~\bibnamefont{Khaykovich}},
  \bibinfo{author}{\bibfnamefont{J.}~\bibnamefont{Cubizolles}},
  \bibinfo{author}{\bibfnamefont{J.}~\bibnamefont{Zhang}},
  \bibinfo{author}{\bibfnamefont{F.}~\bibnamefont{Chevy}},
  \bibinfo{author}{\bibfnamefont{M.}~\bibnamefont{Teichmann}},
  \bibinfo{author}{\bibfnamefont{L.}~\bibnamefont{Tarruell}},
  \bibinfo{author}{\bibfnamefont{S.}~\bibnamefont{Kokkelmans}},
  \bibnamefont{and} \bibinfo{author}{\bibfnamefont{C.}~\bibnamefont{Salomon}},
  \bibinfo{journal}{Phys. Rev. Lett.} \textbf{\bibinfo{volume}{93}},
  \bibinfo{eid}{050401} (\bibinfo{year}{2004}).

\bibitem[{\citenamefont{Shin et~al.}(2008)\citenamefont{Shin, Schunck,
  Schirotzek, and Ketterle}}]{shin2008pd}
\bibinfo{author}{\bibfnamefont{Y.}~\bibnamefont{Shin}},
  \bibinfo{author}{\bibfnamefont{C.}~\bibnamefont{Schunck}},
  \bibinfo{author}{\bibfnamefont{A.}~\bibnamefont{Schirotzek}},
  \bibnamefont{and} \bibinfo{author}{\bibfnamefont{W.}~\bibnamefont{Ketterle}},
  \bibinfo{journal}{Nature} \textbf{\bibinfo{volume}{451}},
  \bibinfo{pages}{689} (\bibinfo{year}{2008}).

\bibitem[{\citenamefont{De~Silva and Mueller}(2006)}]{desilva2006stu}
\bibinfo{author}{\bibfnamefont{T.}~\bibnamefont{De~Silva}} \bibnamefont{and}
  \bibinfo{author}{\bibfnamefont{E.}~\bibnamefont{Mueller}},
  \bibinfo{journal}{Phys. Rev. Lett.} \textbf{\bibinfo{volume}{97}},
  \bibinfo{pages}{070402} (\bibinfo{year}{2006}).

\bibitem[{\citenamefont{Recati et~al.}(2008)\citenamefont{Recati, Lobo, and
  Stringari}}]{recati2008ris}
\bibinfo{author}{\bibfnamefont{A.}~\bibnamefont{Recati}},
  \bibinfo{author}{\bibfnamefont{C.}~\bibnamefont{Lobo}}, \bibnamefont{and}
  \bibinfo{author}{\bibfnamefont{S.}~\bibnamefont{Stringari}},
  \bibinfo{journal}{Phys. Rev. A} \textbf{\bibinfo{volume}{78}},
  \bibinfo{pages}{023633} (\bibinfo{year}{2008}).

\bibitem[{\citenamefont{Vichi and Stringari}(1999)}]{vichi1999coi}
\bibinfo{author}{\bibfnamefont{L.}~\bibnamefont{Vichi}} \bibnamefont{and}
  \bibinfo{author}{\bibfnamefont{S.}~\bibnamefont{Stringari}},
  \bibinfo{journal}{Phys. Rev. A} \textbf{\bibinfo{volume}{60}},
  \bibinfo{pages}{4734} (\bibinfo{year}{1999}).

\bibitem[{\citenamefont{Chevy}(2006{\natexlab{b}})}]{chevy2006dpt}
\bibinfo{author}{\bibfnamefont{F.}~\bibnamefont{Chevy}},
  \bibinfo{journal}{Phys. Rev. Lett.} \textbf{\bibinfo{volume}{96}},
  \bibinfo{pages}{130401} (\bibinfo{year}{2006}{\natexlab{b}}).

\bibitem[{\citenamefont{Amoruso et~al.}(1999)\citenamefont{Amoruso, Meccoli,
  Minguzzi, and Tosi}}]{amoruso1999cet}
\bibinfo{author}{\bibfnamefont{M.}~\bibnamefont{Amoruso}},
  \bibinfo{author}{\bibfnamefont{I.}~\bibnamefont{Meccoli}},
  \bibinfo{author}{\bibfnamefont{A.}~\bibnamefont{Minguzzi}}, \bibnamefont{and}
  \bibinfo{author}{\bibfnamefont{M.}~\bibnamefont{Tosi}},
  \bibinfo{journal}{Eur. Phys. J. D.} \textbf{\bibinfo{volume}{7}},
  \bibinfo{pages}{441} (\bibinfo{year}{1999}).

\bibitem[{\citenamefont{{The damping rate is expected to vanish in the truly
  collisionless limit, a regime difficult to access
  experimentally.}}()}]{dampingrate}
\bibinfo{author}{\bibnamefont{{The damping rate is expected to vanish in the
  truly collisionless limit, a regime difficult to access experimentally.}}}

\bibitem[{\citenamefont{Landau and Lifshitz}(1987)}]{landau1987fmv}
\bibinfo{author}{\bibfnamefont{L.}~\bibnamefont{Landau}} \bibnamefont{and}
  \bibinfo{author}{\bibfnamefont{E.}~\bibnamefont{Lifshitz}},
  \bibinfo{journal}{Course of Theoretical Physics Vol. 6 : Fluid Mechanics}
  (\bibinfo{year}{1987}).

\bibitem[{\citenamefont{Pilati and
  Giorgini}(2008{\natexlab{b}})}]{pilati2008psp}
\bibinfo{author}{\bibfnamefont{S.}~\bibnamefont{Pilati}} \bibnamefont{and}
  \bibinfo{author}{\bibfnamefont{S.}~\bibnamefont{Giorgini}},
  \bibinfo{journal}{Phys. Rev. Lett.} \textbf{\bibinfo{volume}{100}},
  \bibinfo{pages}{30401} (\bibinfo{year}{2008}{\natexlab{b}}).

\bibitem[{\citenamefont{Prokof'ev and
  Svistunov}(2008{\natexlab{b}})}]{prokof'ev08fpb}
\bibinfo{author}{\bibfnamefont{N.}~\bibnamefont{Prokof'ev}} \bibnamefont{and}
  \bibinfo{author}{\bibfnamefont{B.}~\bibnamefont{Svistunov}},
  \bibinfo{journal}{Phys. Rev. B} \textbf{\bibinfo{volume}{77}},
  \bibinfo{eid}{020408} (\bibinfo{year}{2008}{\natexlab{b}}).

\bibitem[{\citenamefont{Bruun et~al.}(2008)\citenamefont{Bruun, Recati,
  Pethick, Smith, and Stringari}}]{bruun2008collisional}
\bibinfo{author}{\bibfnamefont{G.}~\bibnamefont{Bruun}},
  \bibinfo{author}{\bibfnamefont{A.}~\bibnamefont{Recati}},
  \bibinfo{author}{\bibfnamefont{C.}~\bibnamefont{Pethick}},
  \bibinfo{author}{\bibfnamefont{H.}~\bibnamefont{Smith}}, \bibnamefont{and}
  \bibinfo{author}{\bibfnamefont{S.}~\bibnamefont{Stringari}},
  \bibinfo{journal}{Phys. Rev. Lett.} \textbf{\bibinfo{volume}{100}},
  \bibinfo{pages}{240406} (\bibinfo{year}{2008}).

\end{thebibliography}

\end{document}